\documentclass[adp,fleqn]{w-art}
\usepackage{times}
\usepackage{w-thm}
\usepackage[english]{babel}
\usepackage[]{graphicx}

\usepackage{color}
\usepackage{dcolumn}
\usepackage{bm}

\begin{document}

\DOIsuffix{theDOIsuffix}

\Volume{xx}
\Issue{x}
\Copyrightissue{xx}
\Month{07}
\Year{2012}
\pagespan{1}{}

\Receiveddate{15 July 2012}
\Reviseddate{}
\Accepteddate{}
\Dateposted{}


\title{Non-equilibrium Polaritonics - Non-linear Effects and Optical Switching}

\author[R. Frank]{Regine Frank\inst{}%
  \footnote{Corresponding author\quad E-mail:~\textsf{regine.frank@kit.edu}}}

\address[\inst{}]{Institut f\"ur Theoretische Festk\"orperphysik, Karlsruher Institut f\"ur Technologie (KIT), 
Wolfgang - Gaede - Strasse 1, 76131 Karlsruhe, Germany}


\begin{abstract}
In this article a microscopic electronic non-equilibrium effect, highly nonlinear polaritonics, is proposed to
mediate an ultrafast all-optical switching. The electronic band
structure within gold (Au) nano grains shall be modified by external laser
light, namely the Franz-Keldysh effect, and the modified electronic density of
states within the Au grains are coupled to a
single mode photonic waveguide. Using this microscopic polaritonic coupling
without ever including any macroscopic influences due to the geometric
arrangement a strong transmission reduction originating from the established
quantum interference is derived. The
lifetime of the coupled states is heavily dependent on the Fano resonance type
binding and the amplitude of the applied electric field. Besides the Fano
signatures the microscopic
coupling photon-electron-photon leads to a gaped electronic density of
states within the Au nano-grains.  

\end{abstract}

\maketitle 

\section{Introduction}

Plasmonics and polaritonics \cite{Stockman1} may be the most promising candidates in the
race of novel ultrafast technologies which have the potential to reach the
application border \cite{Ctistis, Kirschner, Koos}. Meta-materials 
or plasmonic systems \cite{Gan} are on the border to be engineered for ultrafast electronic switches
\cite{Giessen_nat,Cheng, Azad, Hendry, Chen, Garcia}, and ultrashort laser
science opens the door to their exploration \cite{Stockman, Hommelhoff_2, Muecke}.
However discussing picosecond, femto or attosecond time
scales the framework of equilibrium physics reaches it's limits. When one
speaks about ultrafast non-linear processes in commercially fabricated metallic
or organic structures, impurities naturally play a key role for the efficiency
and the proper operation of such devices. They may either disturb the process or
may play the key role of access to new technologies.
Another fascinating track of physics in the context of ultrafast switches are beyond doubt
quantum-optical functional elements \cite{Hofheinz}.
Yet the combination of both, highly non-linear non-equilibrium physics and quantum
optics is an ansatz which has not been felt out in a broad range up to now.
Solid-state-based elements, in that context, provide a
number of advantages such as scalability, tunable light-matter
interactions and comparably easy handling.
Besides its immense technological importance, 
the theoretical description of externally driven quantum systems 
is a challenge itself,  and vividly discussed
\cite{Freericks, monien, Lubatsch_AdP}.
While the basic underlying equilibrium  physics for 
a system with discrete levels coupled to a
continuum of states is well established \cite{Fano, Liu, Artuso, Singh1, Singh2},
the profound description of light-matter interaction far away from thermodynamical
equilibrium is an active and fascinating area \cite{Frank} of research. 

A very simple model to explore non-equilibrium effects is chosen here in form of a
Fr\"ohlich Hamiltonian which couples classically strong laser radiation to
metal electrons. The modified density of states than interacts with the
waveguide photons by means of a Fano resonance. That coupling is considered to
be weak and therefore can be determined by second order perturbation
theory. The model simply focuses on these very basic procedures and does not
at all claim to be complete, especially geometric finite size effects for high
field intensities are neglected by choosing spherical gold grains. Additionally possible interactions
between the grains by scattered light is not considered here either. The pulse
duration in what follows is chosen such that scattering of electrons within
the metal can be neglected as well. 

\section{Franz-Keldysh Effect and Fano Resonance}

In this article the electronic properties of gold (Au) nano-grains exposed to an external field are
investigated and they are coupled to a single mode photonic silicon-on-insulator hollow core waveguide (SOI). In this non-equilibrium system, the
external time-periodic field generates photo-induced electronic side-bands which can be attributed to
the Franz - Keldysh effect, the AC analogon to the well known Wannier-Stark
effect. Intraband
transitions occurring in every metal within the conduction band feature a more or less small
absorbtion rate, whereas the governing processes are the interband
transitions. In Au a significantly different behavior has been found,
which has been attributed to the electronic specific structure of closed
packed Au, namely to the high polarizability of the $5d^{10}$
cores. The collective resonance exhibits a large red shift to approximately
$2.4 eV$ which leads to the fact that the corresponding intensity is governed
by the interband processes but its existence results from the
occurrence of intraband transitions which can be recognized as step like
structures in the spectrum. These transitions, which result from a non Fermi alike
distribution of states, have been observed by Whetten
{\it et al.}\cite{Whetten}, for Au nano-spheres of diameters below $30 nm$. Their occurrence has been interpreted as the border from bulk like
characteristics to the quantum regime of nano particles. 
The transitions within this single plasmon band provide the basis for the
switching effect. It is shown that the bandstructure of the coupled system of Au
nano-grains and SOI is strongly modified by intense
external laser radiation (Fig.\ref{Fig_4}), which can be tested experimentally
e.g. by two photon photoemission (2PPE). Frequency and amplitude of the external field can be
separately used to modulate the position
in energy of the Floquet sidebands and, therefore sensitively control the generation
of a Fano-resonance with the photonic SOI mode. The calculated lifetimes of
the coupled states prove the extremely fast switching.

\section{Fr\"ohlich Hamiltonian and Floquet-Keldysh Method}

As theoretical setup the Fr\"ohlich Hamiltonian \cite{Mahan, Haug} for
fermion-boson interaction is chosen
which is solved by applying the Keldysh formalism with respect to the
non-equilibrium character of the considered processes on the femto-second time
scale. The full Hamiltonian reads
\begin{eqnarray}
H\!&=&\!\! \sum_{k, \sigma} \! \epsilon_k  c^{\dagger}_{k,\sigma}c^{{\color{white}\dagger}}_{k,\sigma} 
+ \!\! \hbar\omega_o  a^{\dagger}a^{{\color{white}\dagger}}
\!+\!\! g\! \sum_{k, \sigma}
c^{\dagger}_{k,\sigma}c^{{\color{white}\dagger}}_{k,\sigma}  (a^{\dagger}  \! +
a)
\\&&- t\!\! \sum_{\langle ij \rangle, \sigma}\!\!    c^{\dagger}_{i,\sigma}c^{{\color{white}\dagger}}_{j,\sigma}  +\!\! i\vec{d}\cdot\vec{E}_0 \cos(\Omega_L \tau)\sum_{<ij>} 
 \left(
           c^{\dagger}_{i,\sigma}c^{{\color{white}\dagger}}_{j,\sigma} 
 	  -
           c^{\dagger}_{j,\sigma}c^{{\color{white}\dagger}}_{i,\sigma} 	  
 \right).
\end{eqnarray}
The ingredients can be named as the electronic onsite term $\sum_{k,
  \sigma} \! \epsilon_k
c^{\dagger}_{k,\sigma}c^{{\color{white}\dagger}}_{k,\sigma}$, the single
waveguide mode $\hbar\omega_o  a^{\dagger}a^{{\color{white}\dagger}}$,
the perturbative coupling of electrons to waveguide photons (the polariton) $g
\sum_{k, \sigma}
c^{\dagger}_{k,\sigma}c^{{\color{white}\dagger}}_{k,\sigma}  (a^{\dagger} +a)$,
the nearest neighbor hopping of electrons in the lattice $ - t
\sum_{\langle ij \rangle, \sigma}\!\!
c^{\dagger}_{i,\sigma}c^{{\color{white}\dagger}}_{j,\sigma} $, the classical
coupling of the external electrical field amplitude to the dipole moment of
metal electrons and the resulting contribution to the hopping (plasmon) $ i\vec{d}\cdot\vec{E}_0 \cos(\Omega_L \tau)\sum_{<ij>} 
 \left(
           c^{\dagger}_{i,\sigma}c^{{\color{white}\dagger}}_{j,\sigma} 
 	  -
           c^{\dagger}_{j,\sigma}c^{{\color{white}\dagger}}_{i,\sigma} 	  
 \right) $.

This article is organized as follows, considered is (i) the waveguide in 
contact with nano-grains, (ii) the nano-grain in the external field and finally 
(iii) the non-equilibrium solution of the complete system in 
terms of electronic Keldysh - Green's function and waveguide transmission are discussed. The
overall switching is measured as the difference between density of electronic
states of the totally unaffected system compared to the setup including all
effects due to the external laser light and as a consequence the coupling to
the waveguide mode.

\begin{figure}
\begin{center}
  \scalebox{0.7}[0.7]{\includegraphics[clip]{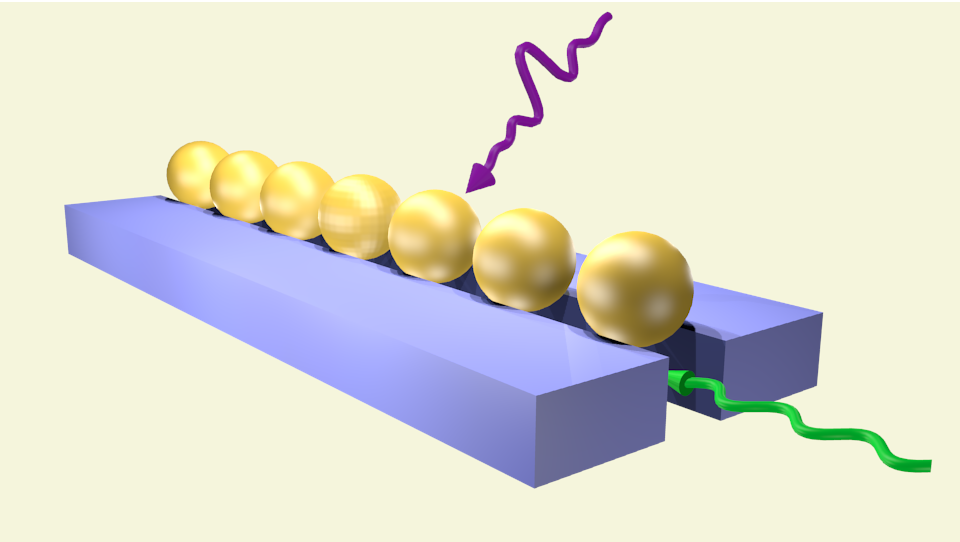}}\end{center}
\caption{Sketched geometry. Gold nanograins in contact with a hollow core SOI
  waveguide. Waveguide photons (green) 
interact with
electrons in the metal nano-grains (Au) forming a coupled light-matter state,
a polariton. This is controlled by an 
external laser (pink), altering the
transmission through the waveguide (green).
}
\label{Fig_00}
\end{figure}

\begin{figure}[t]
\scalebox{1.2}[1.2]{\includegraphics[clip]{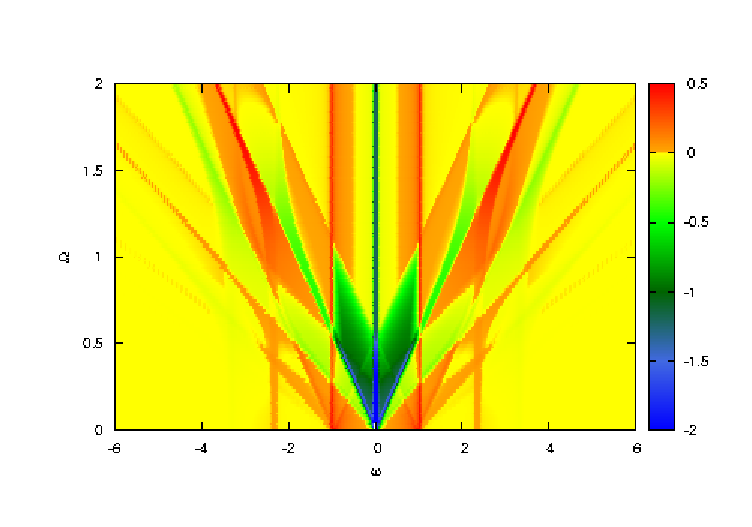}}
\caption{
Laser induced change of the electronic density of
states $\delta G (\omega, \Omega_L) $, Eq.  (\ref{delta_G}). 
The  photonic waveguide mode has  $\hbar\omega_0 \!\!= \!\!2.34eV$, 
the coupling between waveguide photons and electrons is
$g/t\!\!=\!\!0.3$ ($\sim 30\%$ of the coupling $A_0$),
the temperature $T=0$ and the amplitude  $A_0 /t\!\!= \!\!4.0$ (see text).
 $\delta G$ displays a Fano resonance 
around quasiparticle frequencies $\omega\!\!=\!\!\omega_0$, i.e. 
as soon as the external laser field redistributes the  electronic spectral weight such, that
the  waveguide mode finds electrons with about the same
energy to efficiently interact with. 
}
\label{Fig_4}
\end{figure}

As starting point it is chosen a single electron band with nearest-neighbor hopping, 
characterized by the hopping amplitude $t$ , with the dispersion for a cubic
lattice for the ion cores $\epsilon_k= 2
t \sum_i \cos (k_i a)$, $a$ is the lattice constant and $k_i$ are the
components of the wave-vector. A SOI waveguide supporting a single mode
$\hbar\omega_0 \!\!= \!\!2.34eV$. The waveguide itself is
coated and not exposed to the external laser field is assumed, therefore just
metal electrons may interact with the external field and the waveguide is not
affected directly. The electrons
couple with strength $g$ weakly  to
waveguide photons with
 frequency $\omega_0$. A possible setup is depicted in Fig.\ref{Fig_00}.
The Hamiltonian without the external laser field reads, 
\begin{eqnarray}
\label{Hamilton_we}
H\!=\!\! \sum_{k, \sigma} \! \epsilon_k  c^{\dagger}_{k,\sigma}c^{{\color{white}\dagger}}_{k,\sigma} 
+ \hbar\omega_o  a^{\dagger}a^{{\color{white}\dagger}}
\!+g\! \sum_{k, \sigma}
c^{\dagger}_{k,\sigma}c^{{\color{white}\dagger}}_{k,\sigma}  (a^{\dagger}  \! +
a).
\end{eqnarray}
The spatial extension of the Au nano-grains is assumed to be smaller than $ 30
nm$, and consequentially small
compared to the wavelength of the photonic mode inside the
waveguide. Therefore, the momentum of the photons is much smaller than the
electron's momentum and it can be set $q_{\rm photon} \simeq 0$ whenever one considers
the electronic subsystem. Thus, $a^{\dagger}$ ($a$) does not carry an index in
the Hamiltonian. 
In Eq. (\ref{Hamilton_we}), $ \epsilon_k$ is the electronic band energy,
$c^{\dagger}_{k,\sigma}$ ($c^{{\color{white}\dagger}}_{k,\sigma}$) creates
(annihilates) an electron with momentum $k$ and spin $\sigma$. 
$\hbar\omega_0  a^{\dagger}a^{{\color{white}\dagger}}$ is the photon energy eigen-state,
where $a^{\dagger}$ ($a$) creates (annihilates) a photon inside the waveguide
with energy $\hbar\omega_0 $. The last (coupling) term on the r.h.s. is the
standard coupling term between the electronic and the photonic subspaces \cite{Platero}.
Due to the weak interaction of waveguide photons and electrons
inside the nano-grains this interaction is treated perturbatively.
In second order a self-energy contribution is obtained from  Eq. (\ref{Hamilton_we}).
The coupling of the electronic system with a continuous energy spectrum to the
photons with a discrete one leads to a so-called Fano
resonance \cite{Fano}. This is observed in the electronic density of states
(see Fig.\ref{Full}). Here is shown the electron's spectral
function for various frequencies of the waveguide mode for coupling strength
$(\!g/t)^2\!\!=\!\! 0.09$ at zero temperature for a spectral width
$\tau=0.005$ of the
waveguide mode (measured in units of the hopping $t$) at half filling, 
yielding  a suppression of the spectral function around the Fermi-level (half width $\tau$) 
where electrons are transferred to the high (low) energy tails of the spectral
function. Besides the Floquet sidebands the these band structures can be
identified between $\omega=\pm 2.34$ and $\omega=\sim\pm 3.7$ for all
$\Omega$. Their band edges can be  even better identified to be at
$\omega=2.34$ within the laser induced change of the electronic density (see Fig.\ref{Fig_4}). 
It is noted, that if the energy of the waveguide mode $\hbar\omega_0$ is
distinctly different from the energy $\hbar\omega$ of the band-electrons, the
electronic density of states remains approximately unchanged.

Now the subsystem of a nano-grain 
exposed to a
semiclassical electromagnetic laser field is discussed. 
This is described
by the Hamiltonian (Lb = laser + band-electrons) 

\begin{eqnarray}
\label{Hamilton}
H_{Lb}\!=\! - t\!\! \sum_{\langle ij \rangle, \sigma}\!\!    c^{\dagger}_{i,\sigma}c^{{\color{white}\dagger}}_{j,\sigma} 
 \!+\! H_{C}(\tau)
\end{eqnarray}
where $\langle ij \rangle$ implies summation over nearest neighbores.  
$H_{C}(\tau)$ represents the  coupling to the external, time-dependent 
laser field, described by 
the electric field $\vec{E}=\vec{E}_0\cos(\Omega_L \tau)$, via the electronic dipole
operator $\hat{d}(\vec{x})$ with strength $\vec{d}$. It is given by
 \begin{eqnarray}
 \label{def_H_A}
 H_{C} (\tau) &=& i\vec{d}\cdot\vec{E}_0 \cos(\Omega_L \tau)\sum_{<ij>} 
 \left(
           c^{\dagger}_{i,\sigma}c^{{\color{white}\dagger}}_{j,\sigma} 
 	  -
           c^{\dagger}_{j,\sigma}c^{{\color{white}\dagger}}_{i,\sigma} 	  
 \right). 
\end{eqnarray}
The Hamiltonian specifically describes the excitation of a plasmon-polariton,
which correspond to spatially delocalized intraband electronic motion caused by an external
electromagnetic wave. The accelerating energy is
immediately transferred into the motion of electrons by means of single-band
nearest neighbor hopping without interaction between the electrons.
Due to the time dependence of the external field,  Green's functions truly depend
on two separate time arguments. Therefore, a double Fourier
transform from time- to frequency space introducing relative and
center-of-mass frequency is used~\cite{Yeyati_Flores} 
\begin{eqnarray}
\label{Floquet-Fourier}
G_{mn}^{\alpha\beta} (\omega)
\!\! &=&\!\!
\left\lmoustache \!\!{\rm d}{\tau_1^\alpha}\right.\!\!
\left\lmoustache\!\!{\rm d}{\tau_2^\beta}\right.
e^{-i\Omega_L(m{\tau_1^\alpha}-n{\tau_2^\beta})}
e^{i\omega({\tau_1^\alpha}-{\tau_2^\beta})}
G (\tau_1^\alpha,\tau_2^\beta)\nonumber\\
\!\!&\equiv&\!\!
G^{\alpha\beta} (\omega-m\Omega_L, \omega - n\Omega_L),
\end{eqnarray}
where $(m,n)$ label the Floquet modes \cite{Haenggi} and $(\alpha, \beta)$ specify on which branch
of the Keldysh contour ($\pm$) the respective time argument resides. Floquet
states are the fouriertransformed analog to Bloch states: The first
ones result from a time-periodic potential whereas the latter are the
result of a space-periodic potential and both induce a band-structure.
The special case of non-interacting electrons allows an analytical solution for 
$G_{mn}(k,\omega) $ by solving the equation of motion. 
Including photo-induced hopping, the exact retarded Green's function for this sub-system has also been discussed in different context e.g. \cite{Lubatsch_AdP}
\begin{eqnarray}
G_{mn}^{R}(k,\omega) 
=
\sum_{\rho}
\frac
{
J_{\rho-m}\left(A_0\tilde{\epsilon}_k \right)
J_{\rho-n}\left(A_0\tilde{\epsilon}_k \right)
}
{
\omega -\rho\Omega_L - \epsilon_k + i 0^+
}
\end{eqnarray}
where $\tilde{\epsilon}_k$ represents the dispersion relation induced by the
external field Eq. (\ref{def_H_A}) and is different from $\epsilon$ Eq. (\ref{Hamilton_we}).
The $J_n$ are the cylindrical Bessel functions of integer order, 
 $A_0 = \vec{d}\cdot\vec{E}_0 $ and
$\Omega_L$ is the laser frequency. The Bessel function indicate the highly
non-linear characteristics of the switching effect under investigation.
The physical Green's function is given according to
\begin{eqnarray}
\label{EqGsum}
G^{R}_{\rm Lb}(k,\omega)  
=
\sum_{m,n}
G_{mn}^{R}(k,\omega).
\end{eqnarray}

\begin{figure}
\scalebox{1.2}[1.2]{\includegraphics[clip]{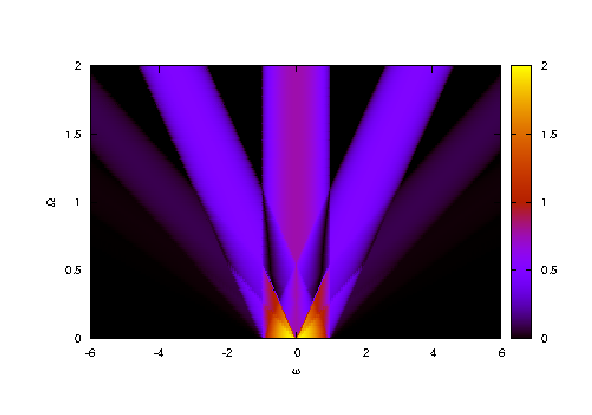}}
\caption{
The imaginary part of the local Green's function, the local density of states
(LDOS) Eq. (\ref {EqGsum}), is depicted as a function of quasiparticle energy,
and frequency respectively
$\omega$ and external laser frequency $\Omega_L$ at zero
temperature for external field amplitude $A_0/t=4.0\,eV$  (see caption text). 
The original semicircular DOS evolves photonic side-bands as the
laser frequency increases. Sidebands of first and second order can be
identified for this specific value of field amplitude.
}
\label{Fig_1b}
\end{figure}

\begin{figure}
\scalebox{1.2}[1.2]{\includegraphics[clip]{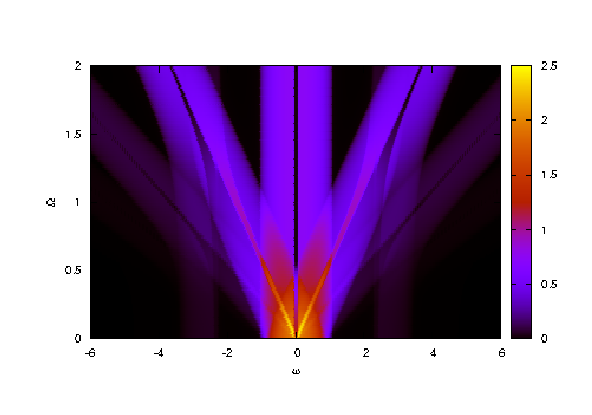}}
\caption{Depicted is the imaginary part of the full Green's function,
  including all interactions as a function of frequency
$\omega$ and external frequency $\Omega_L$ at zero temperature. The external
amplitude is chosen to be $A_0/t=4.0 $  (see text). 
The electronic Floquet-Keldysh (Fig. \ref{Fig_1b}) bandstructure is modified
by the coupling to the single mode waveguide operated at
$\hbar\omega\,=\,2.34\,eV$. The coupling is established by an electron-mediated
Fano resonance. This Fano resonance results in an avoided crossing of bands at
$\hbar\omega\,=\,2.34\,eV$ and $\hbar\Omega\,=\,1.17\,eV$. Spectral weight from the
Fermi edge position in equilibrium is obviously shifted to the center of the
Floquet sidebands by that coupling of
gold and waveguide which results in the development of a gap in the center of
the original band at $\omega\,=\,0.0$. For all $\Omega$ the development of
high (low) energy tails can be observed, which leads to the enhancement
excited bands due to the Fano resonance.
}
\label{Full}
\end{figure}

A numerical evaluation of Eq. (\ref{EqGsum}) in
Fig.\ref{Fig_1b} is presented, where  ${\rm Im}\,G^{R}_{\rm Lb}(k,\omega)$, is displayed as a function of quasiparticle energy
$\hbar\omega$ and external frequency $\Omega_L$ at zero temperature for
$A_0/t=4.0$. 
As a typical value for the hopping $t=1eV$ is chosen. The waveguide is operated
at the frequency $\hbar\omega_0=2.34 eV$ which corresponds to a frequency
doubled Nd-YAG laser ($\hbar\omega_0=1.17 eV$). The external
laser shall be characterized by $10 fs$
pulses, and shall be $E_0=1.008\times 10^{10} V/m$ 
in the surface region of the Au grains, including Mie type \cite{commend}
field enhancement effects due to
the small particle sizes \cite{Plech1, Plech2, Hommelhoff_1}. For the
nano-grains a damage threshold \cite{Boedefeld_Krueger} of $0.5 J/cm^2$ and $|d|=6.528\times 10^{-29}Asm$ ( with the lattice constant for Au $a_{Au}=
4.08 \times 10^{-10}m$) is feasable, 
resulting \cite{Faisal} in $A_0 \equiv d \cdot E_0=4.0 eV $. Furthermore a
waveguide with length $10$ $\mu m$ with a density of $50\%$ nanograins is assumed.
The original semicircular density of states develops
photonic side-bands, the bandstructure, as the external laser frequency
$\Omega_L$ increases. Due to the point-inversion symmetry of the underlying lattice, the first side-band represents here the two-photon processes. The less pronounced second
side-band, therefore, represents four-photon processes.
Their occupation is described by the non-equilibrium distribution function as
calculated from the Keldysh component of the Green's function.

\begin{figure}[t]
\begin{center}\scalebox{0.6}[0.6]{\includegraphics[clip]{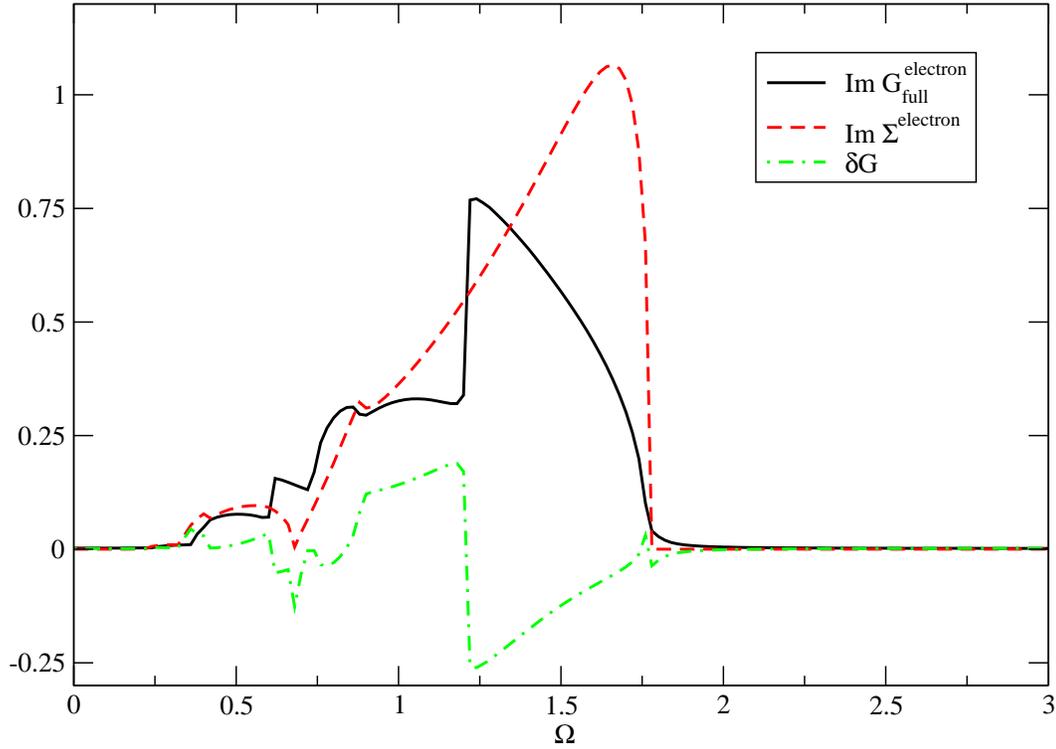}}
\end{center}
\caption{ 
The imaginary part of the retarded component of the full electronic Green's
function $\Im G$ is shown
in combination with the  imaginary part of the retarded component of the
electronic selfenergy $\Im \Sigma$
and the imaginary part of the difference Green's function $\delta G$ as defined
in the text, for a single
electronic energy of $\hbar\omega=2.34$ eV and for a range of external
pump laser frequencies $\Omega$. At a about $\hbar\Omega=1.2$ eV the sharp
dip in $\delta G$ indicates the maximum modulation in the transmittance of the
waveguide mode due to the strong coupling to the nanograins. That point of
switching corresponds t a maximum change of spectral weight in the electronic
density of states $\Im G$. The inverse lifetime $\Im \Sigma$ is found to be
finite valued and the corresponding lifetime of the states equals
$t\,=\,5.908\cdot10^{-15}\,s$. The element can be operated having very short
dead times.
        }
\label{Schnitte_amp4}
\end{figure}

\begin{figure}[t]
\begin{center}\scalebox{0.6}[0.6]{\includegraphics[clip]{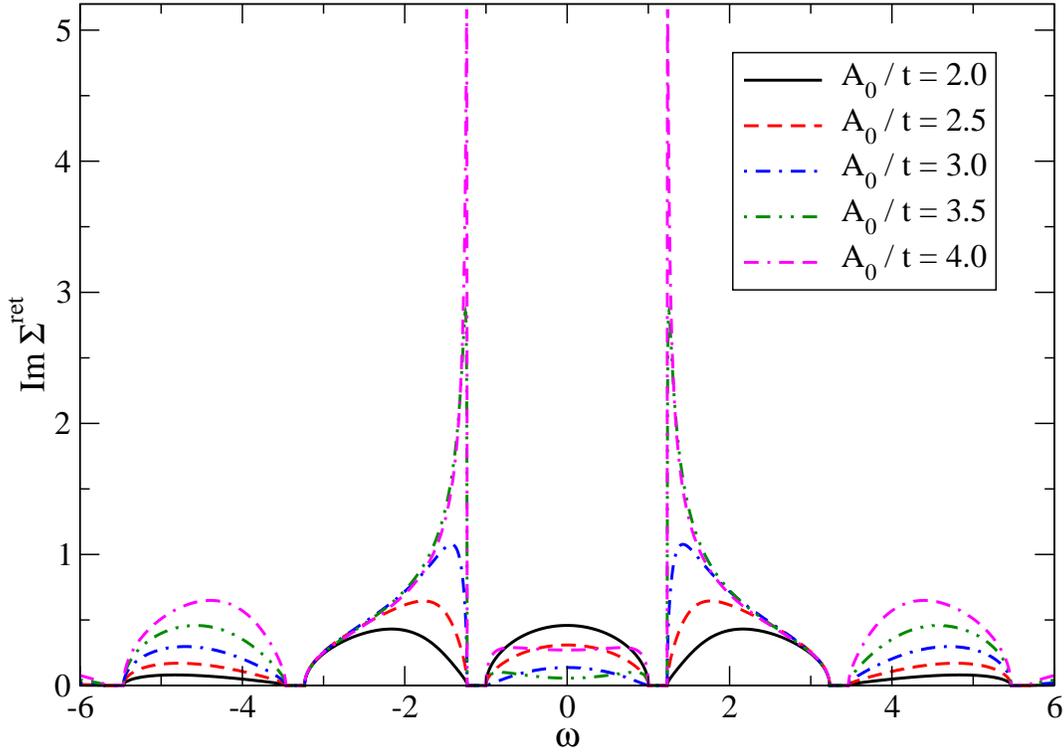}}
\end{center}
\caption{The imaginary part of the retarded component
of the electronic selfenergy $\Im \Sigma$ is shown for the external field
amplitudes  $A_o$ assuming 2.0, 2.5, 3.0, 3.5 and 4.0. $\Im \Sigma$  is
proportional to the inverse lifetime of the electronic excitations. The
development of the inverse lifetime at external field frequency
$\hbar\Omega=1.24 eV$ corresponding to the switching parameters in the text is
displayed. It can be found that the inverse lifetime for increasing external
field amplitude drastically increases. The liftime therefor decays rapidly
(see text). The change of the lifetime of states in experiments indicates
non-equilibrium behavior.}
\label{Composed_Pic}
\end{figure}


\begin{figure}[t]
\begin{center} \scalebox{0.55}[0.55]{\includegraphics[clip]{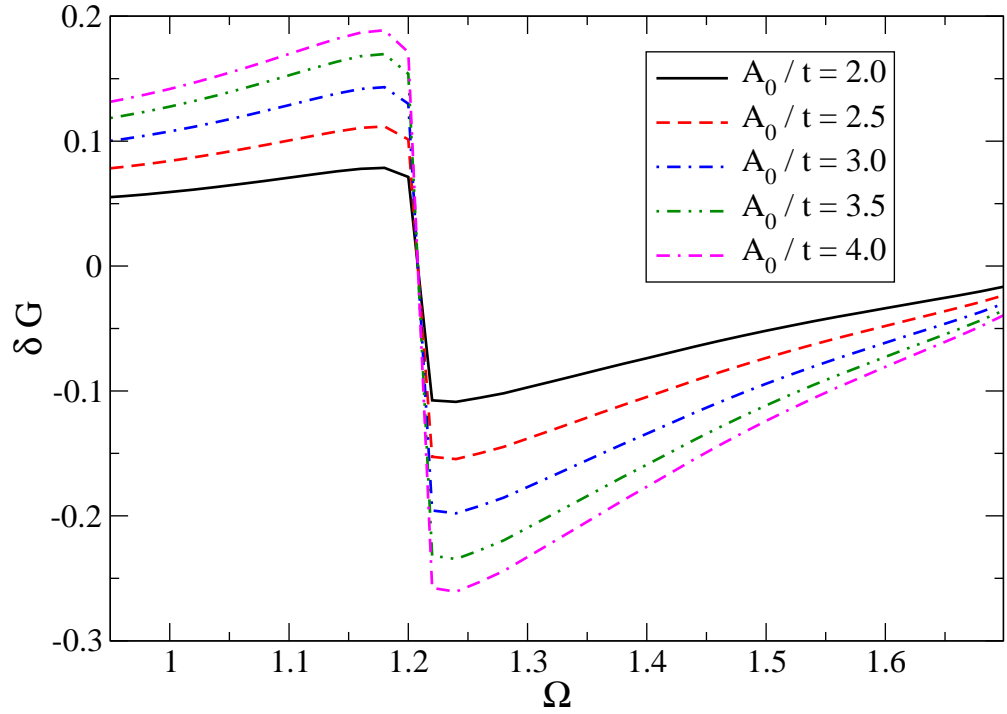}}\end{center}
\caption{
The relative switch of electronic density of states $\delta G$ is a measure for
the dip in the transmission of waveguide photons and therefore the inverse
polaritonic coupling between waveguide photons and electrons. The waveguide is of unit length $l\!\!=\!\!l_o$. The graph is displayed as a function of 
external laser frequency $\Omega_L$,  for four different amplitudes of the
external field. Parameters are given in the text.
}
\label{Fig_m}
\end{figure}

\begin{figure}[t]
\begin{center} \scalebox{0.55}[0.55]{\includegraphics[clip]{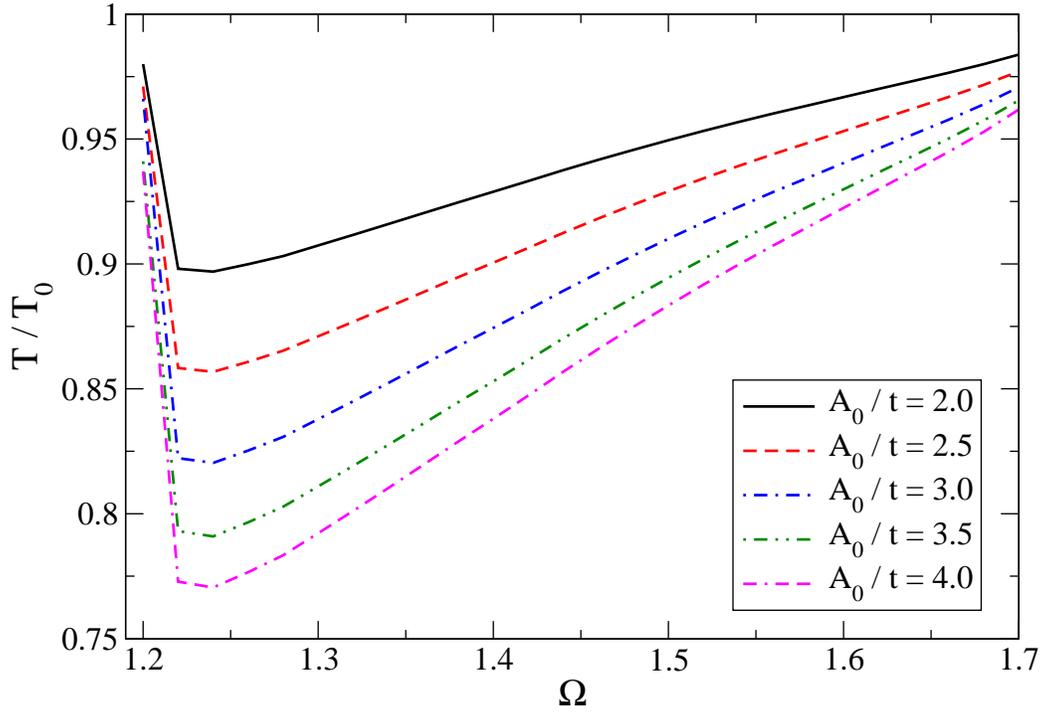}}\end{center}
\caption{
Relative photon transmission 
in a waveguide of unit length $l\!\!=\!\!l_o$ as a function of 
laser frequency $\Omega_L$,  for four different amplitudes of the external field.
Parameters are given in the text.
}
\label{Fig_m}
\end{figure}

In a last step, the relaxation processes due to the interaction
between the band electrons and the waveguide as described by
Eq. (\ref{Hamilton_we}) are combined, with the impinging external laser as introduced
in Eq. (\ref{Hamilton}).
The resulting Green's function consequentially describes the photonic waveguide
with Au nano-grains that themselves are now exposed to the external laser
radiation. In this non-equilibrium system the weak coupling between the electrons and the waveguide
photons is treated by second order perturbation theory and the interaction between the
electrons and external laser in terms of the Floquet theory, as has been shown above.

Since the focus lies on possible switching effects, the
initial situation is chosen such that the photonic mode, $\hbar\omega_0 = 2.34 eV$,  
is far off the electronic bandedge. Electrons arrive
band energies ranging from $-1\le\hbar\omega/t\le+1$.
A Fano resonance is only observable in the presence of external radiation
of appropriate frequency, i.e. only if the induced side-bands meet 
the energy of the waveguide mode.
This resonance will also affect the transmission properties in the waveguide,
since now waveguide photons can be absorbed in the formation of a
mixed state of light and matter with the laser-induced charge excitations in the
Au nano-grains. Thus a waveguide polariton is created yielding to a
significant reduction of the waveguide's transmission.
In \ref{Fig_4}, the laser induced change of the density of states $\delta G$ as a function
of quasi-particle energy $\hbar \omega$ and external laser frequency
$\Omega_L$ is displayed. 
The quantity $\delta G $ defined according to
\begin{eqnarray}
\label{delta_G}
\delta G 
&=&
\left[ 
{\rm Im\,} G_{\rm } (\omega, \Omega_L)
-
{\rm Im\,} G_{\rm Lb} (\omega, \Omega_L)
\right]\\
&&-
\left[ 
{\rm Im\,} G_{\rm wb} (\omega)
-
{\rm Im\,} G_{\rm b} (\omega)
\right]\nonumber
\end{eqnarray}
measures the effect of the impinging laser field on the electronic density of
states, and vanishes as the external laser and the coupling to the
waveguide is turned off. In Eq. (\ref{delta_G}),  $G_{\rm }$ represents
the Green's function including all processes, $G_{\rm Lb}$ the interaction
between the laser field and the band electrons as given in
Eq. (\ref{EqGsum}), $G_{\rm wb}$ describes the waveguide in presence of the
band electrons and is solution to Eq. (\ref{Hamilton_we}), and finally 
$ G_{\rm b}$ is the Green's function of just the non-interacting band electrons.
In \ref{Full} the laser induced electronic density of
states coupled to the waveguide $G (\omega, \Omega_L) $ experiences a Fano resonance as
soon as the external laser redistributes electronic spectral weight such, that
the  waveguide mode at $\hbar \omega_0 = 2.34 eV$ 
may efficiently be absorbed. This is derived when the first photonic
side-band meets the energy of the photonic waveguide
mode yielding a sign change in $\delta G$ at this energy; compare also to Fig.\ref{Fig_1b}.
Additionally, one finds in Fig.\ref{Full} at this particular 
point an avoided crossing of the bands.
In Fig.\ref{Fig_4}, the laser induced change of the electronic
density  of states $\delta G $ is shown at fixed quasiparticle energy $\hbar
\omega=\hbar \omega_0$,
where $\omega_0$ is the frequency of the photonic waveguide mode. 
Asymptotically, i.e., for large $\Omega_L$, $\delta G$
vanishes, as already indicated in \ref{Fig_1b}, because in this limit
there is no electronic spectral weight at the particular energy of the waveguide mode.
In the opposite limit, $\Omega_L \rightarrow 0$, the influence of the laser
field is non-zero, because here higher-order laser induced side-bands exist,
yielding spectral weight at the resonance position already for smaller laser
frequencies. That result can also be concluded from the second-order side-band in
\ref{Full}.

In a waveguide of length $l$, the ratio between the initial and the transmitted intensity is given by $T\sim \exp(-\alpha\, l/l_o)$.
Here $\alpha / l_o$,  is the absorption coefficient divided by the unit length
$l_o$, where $\alpha$ includes an average over one period of the external
periodic driving field with frequency $\Omega_L$. 
It can be recognized that $\omega\delta G $ can be understood as the
leading contribution to the relative absorption coefficient as discussed in
detail in
ref. \cite{Jauho_1}. 
In \ref{Fig_m} is shown the relative transmission of photons
$T/T_0$
within the waveguide of unit length $l=l_o$ as a function of the external
laser frequency $\Omega_L$. 
Depending on the frequency of the driving field, an intensity drop of up to
$25 \%$ is observed. By increasing the length of the waveguide or the density
of the grains this effect is enhanced. Therefore one can conclude, that by switching 
laser light of an appropriate frequency of about $\hbar \Omega_L/t \sim 1.35 $, the transmitted photon intensity through the waveguide 
can be significantly altered, and by varying the length of the waveguide the
transmission inside the waveguide can in fact be turned on and off.  The calculated lifetime of the
coupled state in the position of the Fano resonance for the presented
parameters equals $t\,=\,5.908\cdot 10^{-15}\,s $. After that reaction time
the switch should be in the initial state and ready for the next signal. In
Fig.\ref{Composed_Pic} the evolution of the 
corresponding lifetimes with increasing external laser
amplitude is discussed. It is found, that the Floquet sidebands evolve with raising field
amplitude and so does the calculated inverse lifetime. The inverse lifetime of
coupled states with raising field
amplitude is diminished. It can be clearly observed, that the minimum is
shifting with increasing amplitude towards the position of the lower band
edges for $\omega>0$ and to the upper band edges for $\omega<0$. 

\section{Conclusion}

To conclude: In this article a quantum field theoretical model
for a photonic waveguide in contact with gold nano-grains
which themselves are exposed to external laser irradiation has been presented.
The strong and coherent external laser is  described in terms of 
the Floquet theory, assuming classical behavior of this oscillatory-in-time
field, whereas the interaction with the waveguide mode reflects a quantum
interference. The  obtained results demonstrate the high potential  of
waveguide polaritons for all-optical switching. Both, frequency and amplitude of the external laser control transmission through the waveguide, 
and each of these features ensure ultrafast switching processes.

Acknowledgments: The author wants to thank F. Hasselbach, P. Hommelhoff, A. Lubatsch,
W. Nisch, G. Sch\"on and M. Wegener for constructive discussions. The author
is fellow of Karlsruhe School of Optics $\&$ Photonics (KSOP).\\


\begin{thebibliography}{9999999}
\bibitem{commend}
Spacially resolved selfconsistent calculations showing the unisotropic Mie type field
amplification coupled to the non-equilibrium calculation of the density of
states will be presented separately.
\end{thebibliography}
\end{document}